\documentclass{iaus}
\usepackage{graphicx}

\title[Measuring Warps with Spectroscopy] 
{Measuring Outer Disk Warps with Optical Spectroscopy}

\author[Daniel Christlein \& Joss Bland-Hawthorn]   
{Daniel Christlein$^1$
 \and Joss Bland-Hawthorn$^2$}

\affiliation{$^1$Max-Planck-Institut f\"ur Astrophysik, \\
Karl-Schwarzschild-Str. 1, 85748 Garching, Germany \\ email: {\tt dchristl@mpa-garching.mpg.de} \\[\affilskip]
$^2$Institute of Astronomy, School of Physics \\ University of Sydney,
NSW 2006, Australia \\email: {\tt jbh@physics.usyd.edu.au }}

\pubyear{2008}
\volume{254}  
\pagerange{119--126}
\jname{The Galaxy Disk in a Cosmological Context}
\editors{J. Andersen, J. Bland-Hawthorn \& B. Nordstr\"om, eds.}
\begin{document}

\maketitle

\begin{abstract}
Warps in the outer gaseous disks of galaxies are a ubiquitous
phenomenon, but it is unclear what generates them. One theory is
that warps are generated internally through spontaneous bending
instabilities. Other theories suggest that they result from the
interaction of the outer disk with accreting extragalactic material. In this case, we expect to find cases
where the circular velocity of the warp gas is poorly correlated with
the rotational velocity of the galaxy disk at the same radius. Optical spectroscopy presents itself as an interesting
alternative to 21-cm observations for testing this prediction, because (i)
separating the kinematics of the warp from those of the disk requires
a spatial resolution that is higher than what is achieved at 21 cm at
low HI column density; (ii) optical spectroscopy also provides
important information on star formation rates, gas excitation, and
chemical abundances, which provide clues to the origin of the gas in warps. We present here preliminary results of a study of the kinematics of gas in the outer-disk warps of seven edge-on galaxies, using multi-hour VLT/FORS2 spectroscopy.
\keywords{galaxies: evolution, galaxies: formation, galaxies:
kinematics and dynamics, galaxies: structure}
\end{abstract}

\firstsection 
\section{Introduction}

Warps in the outer disks of galaxies are a ubiquitous phenomenon. They
are seen both in the distribution of stars \cite[(Sanchez-Saavedra et
al. 1990; Cox et al. 1996)]{sanchezsaavedra,cox} and neutral hydrogen \cite[(e.g., Sancisi
1976; Bosma 1981)]{sancisi,bosma}, and surveys estimate that
possibly $50\%$ or more of all galaxies show evidence for warps beyond the isophotal radius R$_{25}$ \cite[(Briggs 1990)]{briggs}.
(which typically contains $\sim 90\%$ of the total light).
 Usually, these manifest themselves in the form of the outer disk bending away from the plane defined by
the inner disk, defining either the shape of a bowl, or, more
commonly, an integral sign as seen edge-on. This suggests that
specific angular momentum of material in the warps is not aligned with
that of the inner disk.

Numerous suggestions have been made over the years for the responsible
mechanisms. Among the earliest
such proposals were internal bending modes in the disk
\cite[(Lynden-Bell 1965)]{lyndenbell}, but such modes were soon
recognized to be persistent only in a disk with an unrealistically
sharp mass truncation \cite[(Hunter \& Toomre
1969)]{huntertoomre}. \cite[Revaz \& Pfenniger (2004)]{revaz} have revived this
discussion by identifiying short-lived bending instabilities as a
possible cause. Other proposed mechanisms focus on the interaction of
the baryonic disk with its environment: A bending of the outer
baryonic disk may be introduced by a misalignment between the angular
momentum of the inner baryonic disk and the hypothetical
non-spherical, dark matter halo that it is embedded in \cite[(Toomre
1983; Dekel \& Shlosman 1983; Kuijken 1991; Sparke \& Casertano
1998)]{toomre,dekel,kuijken,sparke}, creating a gravitational torque
on the disk. However, it has been argued that
the inner halo would realign with the baronic disk over time, and the
warp would dissipate \cite[(Nelson \& Tremaine 1999; Binney, Jiang \&
Dutta 1998; Dubinski \& Kuijken 1995; New et al. 1998)]{nelson,bjd,dubinski,new}.
 \cite[Ostriker \& Binney (1989), Jiang \& Binney (1999), Shen \&
Sellwood (2006)]{ostriker,hiang,shen} consider the impact
of ongoing accretion onto an outer dark matter halo and argue that,
since the angular momentum of infalling material will in general not
be aligned with the present galaxy disk, this will create an ongoing
torque on the outer galaxy disk. \cite[Binney (1992)]{binney91} has
furthermore suggested that, if infalling material loses angular
momentum to the halo, it might penetrate as far as the outer edge of
the baryonic disk itself.

It is this hypothesis that we wish to test with the present
work. How galaxies acquire gas is one of the key questions in our
understanding of how they evolve,  and determining whether warps are
indeed signatures of such processes therefore is of great importance. 

How can the direct accretion hypothesis be tested? The specific
angular momentum vector of accreting material will, in general, neither be aligned exactly with that of the inner disk,
nor have the same size. If such infalling material is indeed in
direct contact and exchanging angular momentum with gas in the
outermost baryonic disk, then it will introduce kinematic anomalies,
i.e., a deviation from disk-like rotation, such as a lag or excess in the circular velocity. Measuring the
circular velocity of gas in the warps therefore becomes an important
observational discriminator.

Our project has measured line-of-sight velocities of gas in the outer
disks of seven galaxies, the majority of which display clear signs of
optical warps. Our measurements were obtained via optical spectroscopy
of the H$\alpha$ line. Although the tradititional way of observing gas
in the outer disk is via the 21-cm line of neutral hydrogen, optical
spectroscopy has proven a surprisingly successful alternative for
studying the outer disk \cite[(Bland-Hawthorn, Freeman \& Quinn 1997;Christlein \& Zaritsky 2008)]{bhfq,cz}, for a number of reasons: 1) Most importantly,
the spatial resolution --- an order of magnitude better than even the
highest-quality interferometric HI maps --- allows us to clearly
separate gas in the warps from gas in the plane of the galaxy. This,
in turn, allows us to access galaxies with smaller angular diameters
at larger redshifts, greatly increasing the number of suitable
targets. 2) Sporadic local star formation or illumination of gas in
the warps by escaping UV flux from the inner star-forming disk
guarantee low-level H$\alpha$ flux from the outer disk far beyond what
is usually perceived as the star formation threshold. It has been demonstrated \cite[(Bland-Hawthorn, Freeman \& Quinn 1997;Christlein \& Zaritsky 2008)]{bhfq,cz} that multi-hour
optical spectroscopy of such low-level emission can, in some cases, probe outer galaxy disks to
similar extents as 21-cm. 3) Optical spectroscopy yields a plethora of
ancillary data, particularly stellar continuum and metal lines. In
determining the origin of gas in the outer disks and warps,
metallicity indicators may provide valuable additional insight
(although an analysis exceeds the scope of these proceedings). Our aim
is to a) measure the rotational velocity of gas in the warps and b)
determine whether there are kinematic anomalies in this rotational
velocity, i.e., sudden breaks or upturns that are not consistent with
an extrapolation of the rotation curve from the inner disk, which may
be associated with the onset of the warp.

\section{Observations}

Our sample consists of seven galaxies, of which six were taken from
the catalog of optically warped galaxies by
\cite[Sanchez-Saavedra et al. (2004)]{sanchezsaavedra2}; one object was targetted blindly without
prior knowledge of a warp. All are nearly edge-on, with
redshifts in the range of several thousand km/s, and their angular size is well-matched to the field of
view.

Our observations were carried out over three nights in September 2007,
using FORS2 on the VLT-UT1. The typical observing strategy was to
observe for a total of one hour with a long slit of
0.5$^{\prime\prime}$ width aligned along the major
axis (to obtain a reference for the rotational velocity and angular
extent gas in the plane of the galaxy), then for two hours with a
position angle offset from the major axis PA by a few degrees. The
purpose of the latter observation was to assure that H$\alpha$ emission
emanating from the warps would be observed, rather than from the plane of the galaxy. Sense
and size of this offset were determined on a case-by-case basis after
inspecting plates from the Digitzed Sky Survey as well as our own
acquisition images, and typically chosen so that the slit would pass
within or just beyond the outermost contours of stellar continuum
light at the end of the warp.

\section{Results}

\begin{itemize}
\item {\bf ESO 184-G063} (Fig. \ref{fig_eso184}) is a small ($M=-18.5$) Sb-type galaxy at cz = 3207
km/s, and has a strong integral-sign warp, which is distinctly
stronger on one side than the other. 

This figure, as well as all subsequent ones, shows the rotation curve
along the major axis (top panel) and in the offset position (bottom
panel; major axis rotation curve plotted with small dots/dashed lines for reference). The radial extent is 2
R$_{25}$ on both sides of the nucleus; the acquisition image is
plotted to the same scale.

Along the major axis, our rotation curve extends to 1.4
R$_{25}$, which in itself is remarkable, given that normal H$\alpha$
rotation curves with conventional exposure times and smaller
instruments typically reach $0.7 R_{25}$, and rarely as far as
R$_{25}$. Our off-axis spectrum extends out to 1.6 R$_{25}$, clearly
intersecting the tip of the optical warp. Both the major axis and the
offset rotation curves are consistent, and there is no sign of a
sudden break in the rotational velocity coincident with the onset of
the warp.

\begin{figure}
\begin{minipage}[t]{0.5\linewidth} 
\centering
\includegraphics[width=6cm]{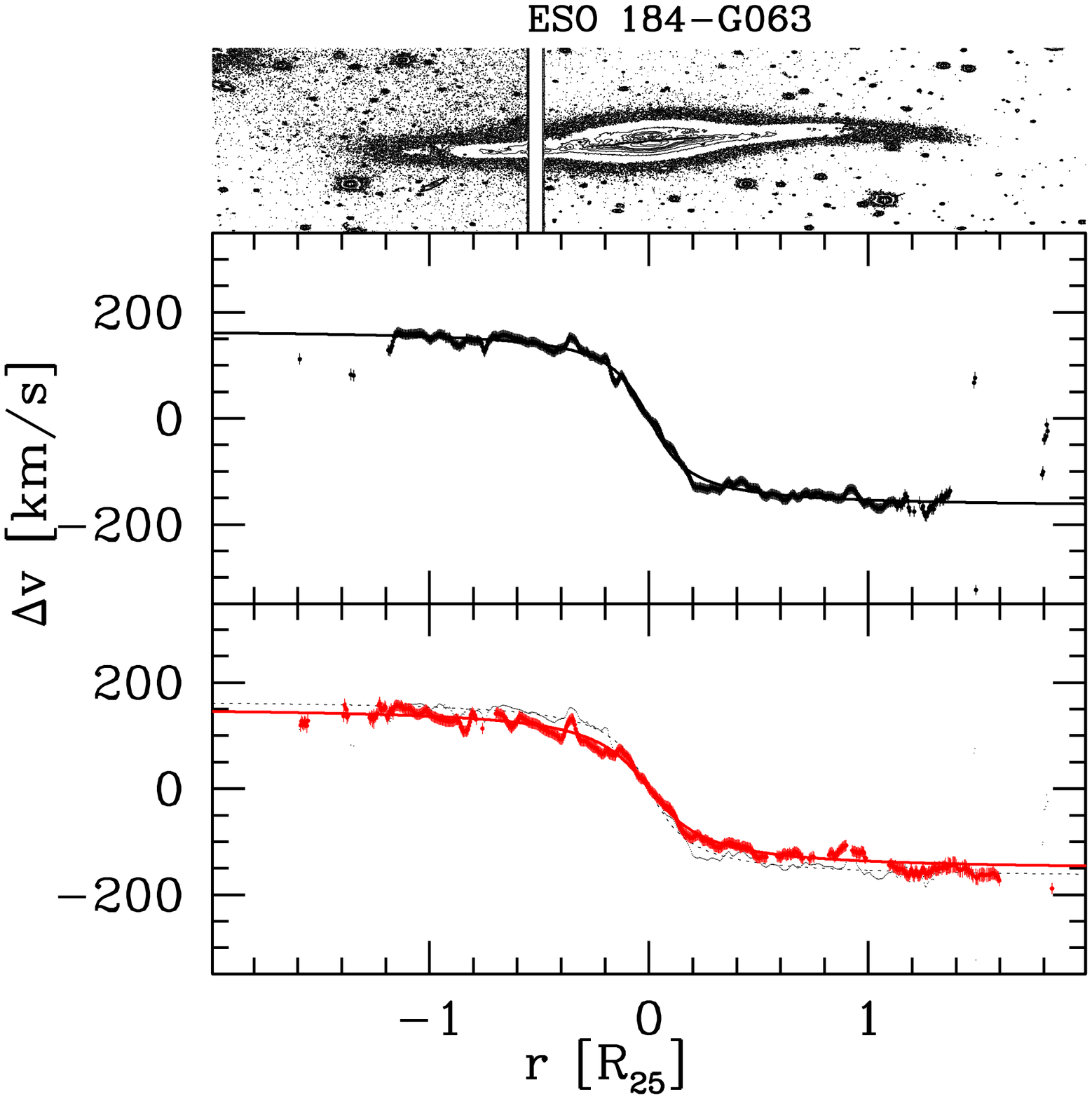}
\caption{ESO 184-G063. For explanation, see text. H$\alpha$ is more
extended in the offset spectrum than along the major axis, but no
kinematic disruption is associated with the warp.}
\label{fig_eso184}
\end{minipage}
\hspace{0.5cm} 
\begin{minipage}[t]{0.5\linewidth}
\centering
\includegraphics[width=6cm]{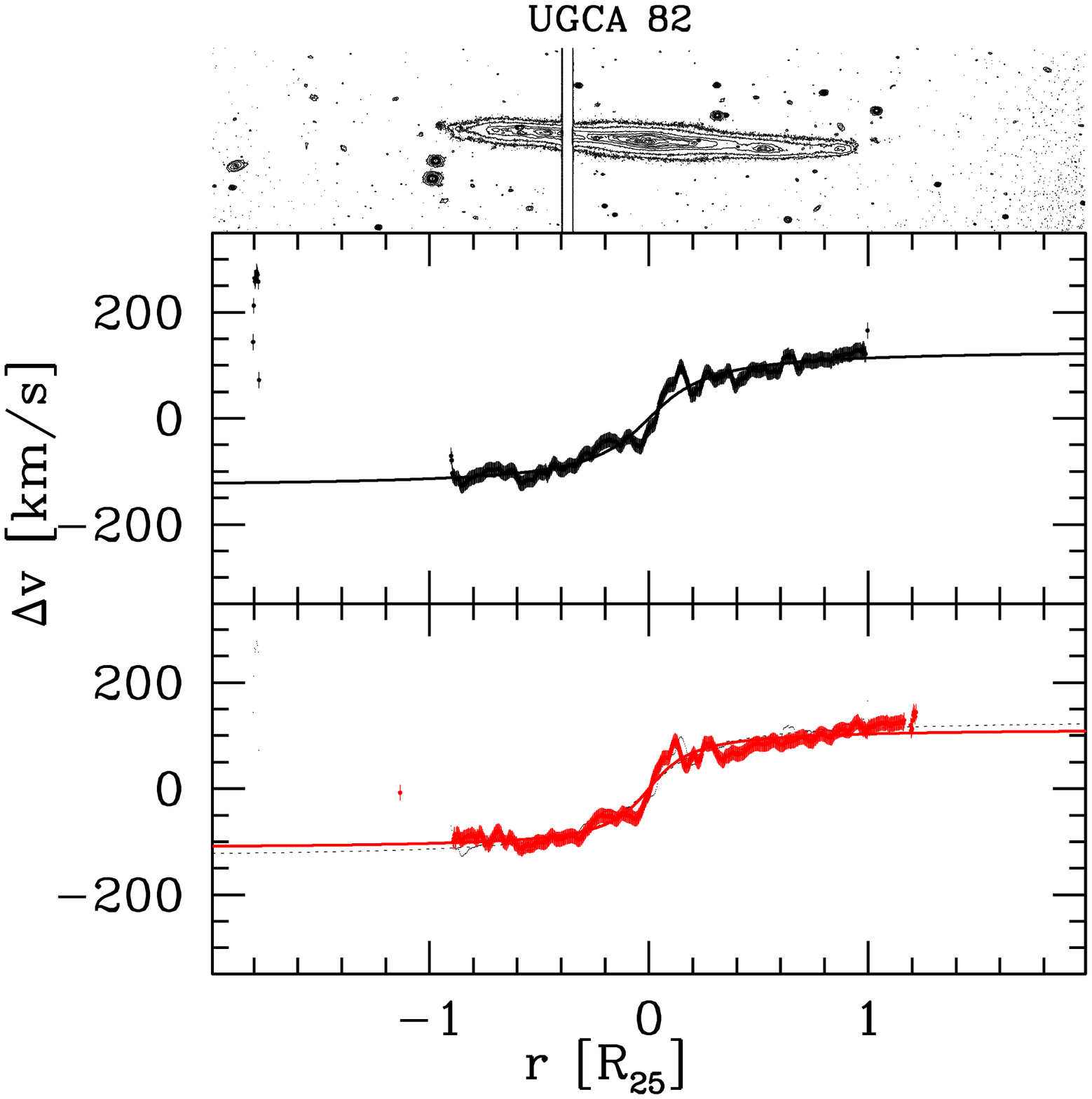}
\caption{UGCA 23. For explanation, see text. H$\alpha$ is more
extended in the offset spectrum than along the major axis, but no
kinematic disruption is associated with the warp. }
\label{fig_ugca23}
\end{minipage}
\end{figure}

\item
{\bf UGCA 23} (Fig. \ref{fig_ugca23}) is an $M=-19$ Sd-type galaxy at cz=3864 km/s. Acquisition
images show lumpy emission and slightly bent isophotes on one
side. The major axis rotation curve extends to $\sim R_{25}$ in this
case. The offset spectrum samples the warp feature and shows emission
as far as 1.2 R$_{25}$. The two rotation curves are indistinguishable;
the kinematics of the extraplanar feature are disk-like.

\item 
{\bf MCG -01-10-035} (Fig. \ref{fig_mcg101}) is a small ($M=-18.3$) Sc-type galaxy at cz = 3788
km/s. Our acquisition images reveal a very extended tail of low
surface brightness extending as far as $2R_{25}$ and then curving
back. Its appearance is very suggestive of a tidal feature. While the
other side of the galaxy does not exhibit an equivalently striking
feature, the outer isophotes appear bent on both sides.

The major axis spectrum samples far into this low-surface brightness
tail and extends to 1.6 R$_{25}$ on that side, and $1.4 R_{25}$ on the
other. Remarkably, at the onset of this feature, around R$_{25}$, we
see a sudden break in the rotation curve {\it along the major axis} in
the sense of a jump to lower rotational velocities by $\sim 60-70$
km/s. On the other side, no such strong break is discernible, but
there is H$\alpha$ emission at velocities distinctly below the
extrapolated rotation velocity, going as far down as the systemic
velocity of the galaxy itself. 

Our offset spectrum intercepts the tail at its largest extent. While
we do not see a continuous H$\alpha$ signal beyond R$_{25}$, there are
individual H$\alpha$-bright regions extending as far as $2 R_{25}$ on
both sides of the galaxy. The rotation curve fitted to this spectrum
shows no breaks in the velocity; however, at large radii, its
line-of-sight velocity is consistent with the lower-velocity gas in the tidal feature.

\begin{figure}
\begin{minipage}[t]{0.5\linewidth} 
\centering
\includegraphics[width=6cm]{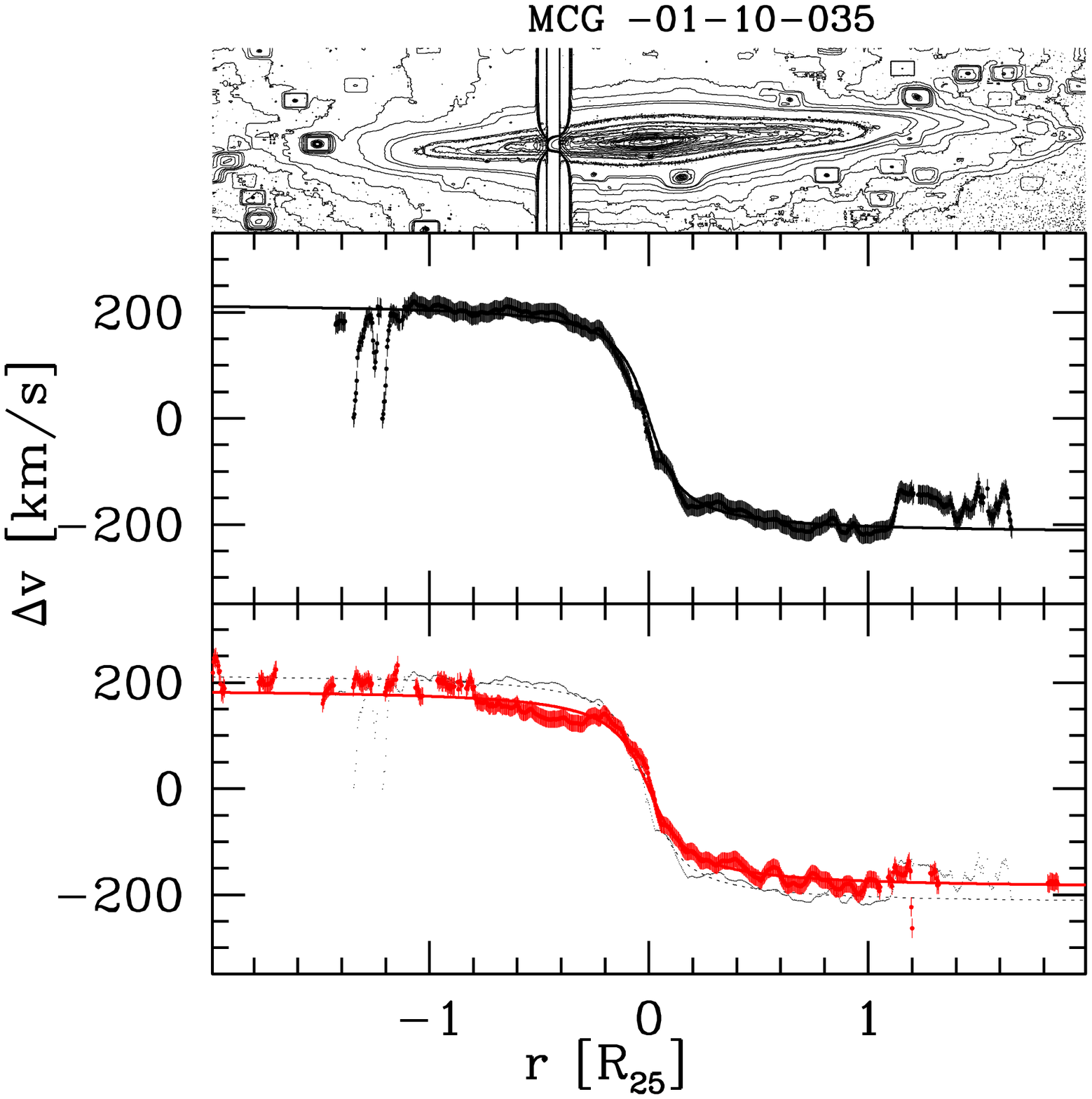}
\caption{MCG -01-10-035. H$\alpha$ is more
extended in the offset spectrum than along the major axis. A break in
the rotation curve is found along the major axis, coinciding with the
onset of a tidal-tail-like warp feature.}
\label{fig_mcg101}
\end{minipage}
\hspace{0.5cm} 
\begin{minipage}[t]{0.5\linewidth}
\centering
\includegraphics[width=6cm]{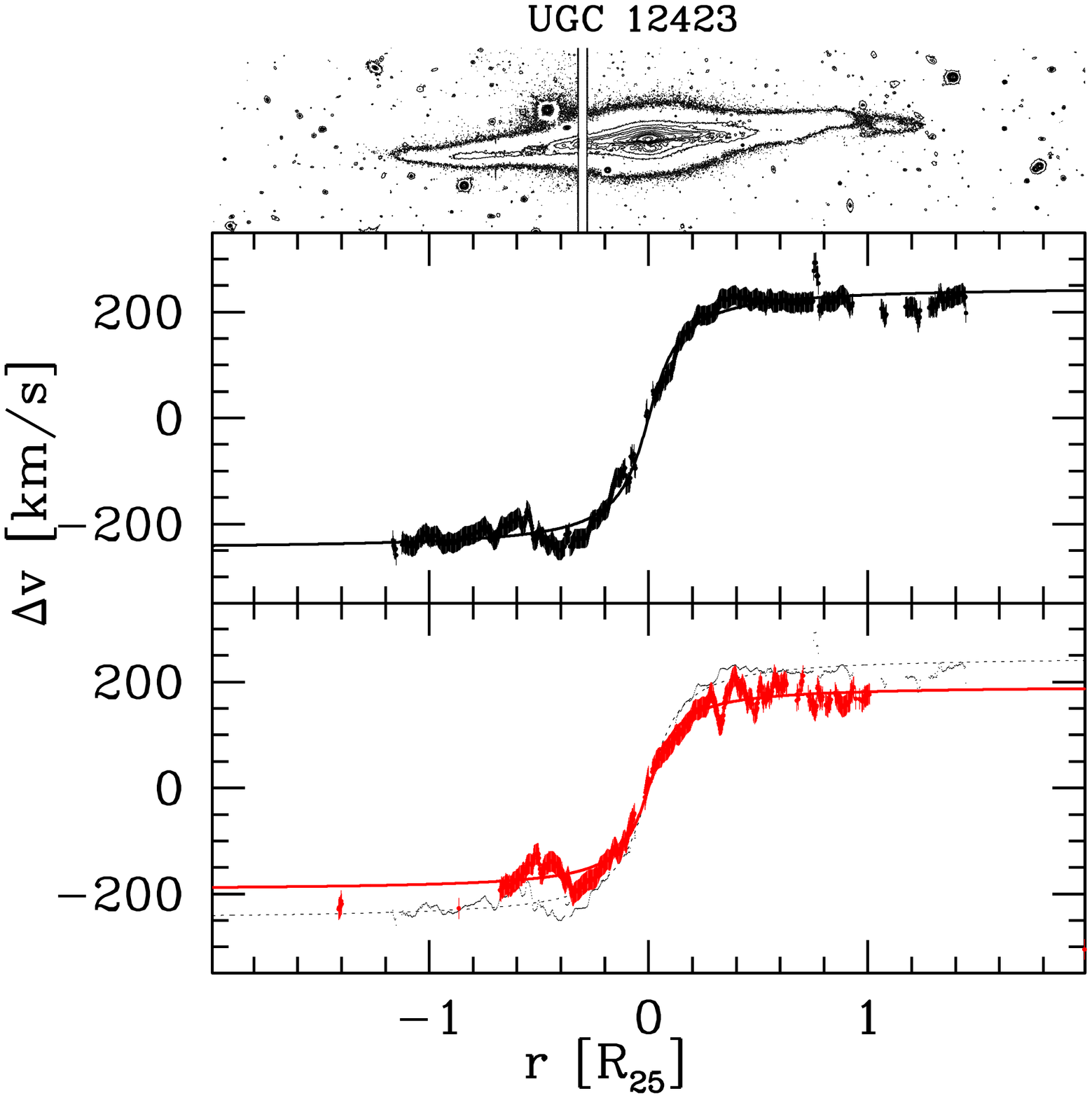}
\caption{UGC 12423. See text for explanation.}
\label{fig_ugc12423}
\end{minipage}
\end{figure}

\item
{\bf UGC 12423} (Fig. \ref{fig_ugc12423}) is an Sc-type galaxy at cz = 4839 km/s with an absolute
magnitude $M=-19.4$. It shows a strong and asymmetrical optical
warp. The major-axis rotation curve extends to 1.4 R$_{25}$ on the
side with the stronger warp, which is even farther than the extent of
the warped disk. At the outermost point, the rotational velocity is
still consistent with the extrapolated rotation curve. A possible dip
in the rotation velocity at the position of the warp feature does not
appear significant.

The offset spectrum is, in this case, much less extended; it reaches
R$_{25}$ on the side of the stronger warp, and only 0.7 R$_{25}$ on
the other side, with the exception of one small H$\alpha$-emitting
region at 1.4 R$_{25}$. There are no signs for any kinematic anomalies
in this rotation curve.


\item {\bf NGC 259} (Fig. \ref{fig_ngc259}) is among the brighter galaxies in this sample ($M=-20$), an
Sbc-type at cz=4045 km/s. It is also the least inclined among our
targets. The acquisition images show that the supposed warp feature is
in fact more suggestive of a continuation of a spiral arm. The major
axis spectrum samples its kinematics out to 1.2 R$_{25}$, but shows no
kinematic disturbance. The offset spectrum only extends to $\sim
R_{25}$.

\begin{figure}
\begin{minipage}[t]{0.5\linewidth} 
\centering
\includegraphics[width=6cm]{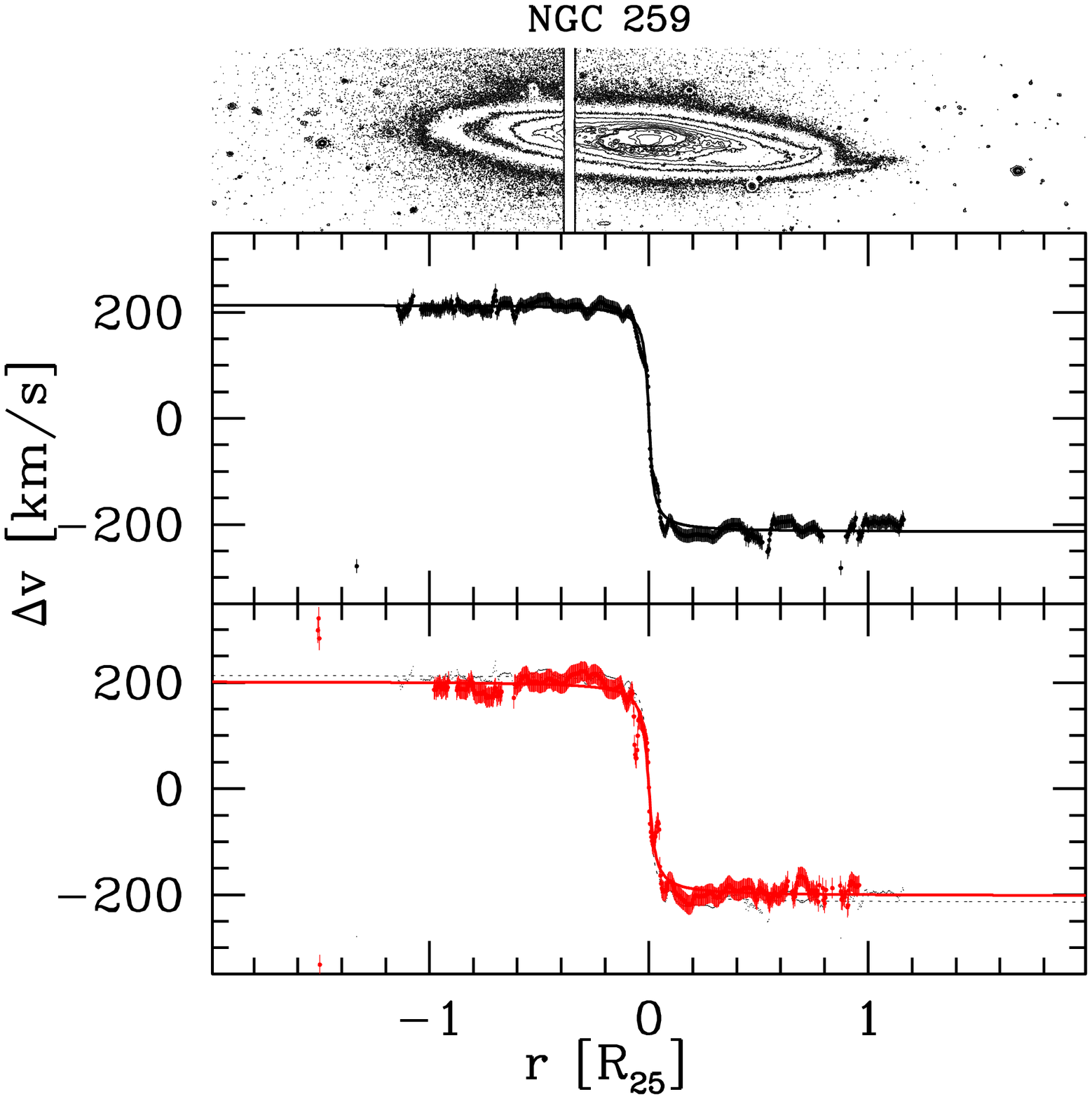}
\caption{NGC 259. See text for explanation.}
\label{fig_ngc259}
\end{minipage}
\hspace{0.5cm} 
\begin{minipage}[t]{0.5\linewidth}
\centering
\includegraphics[width=6cm]{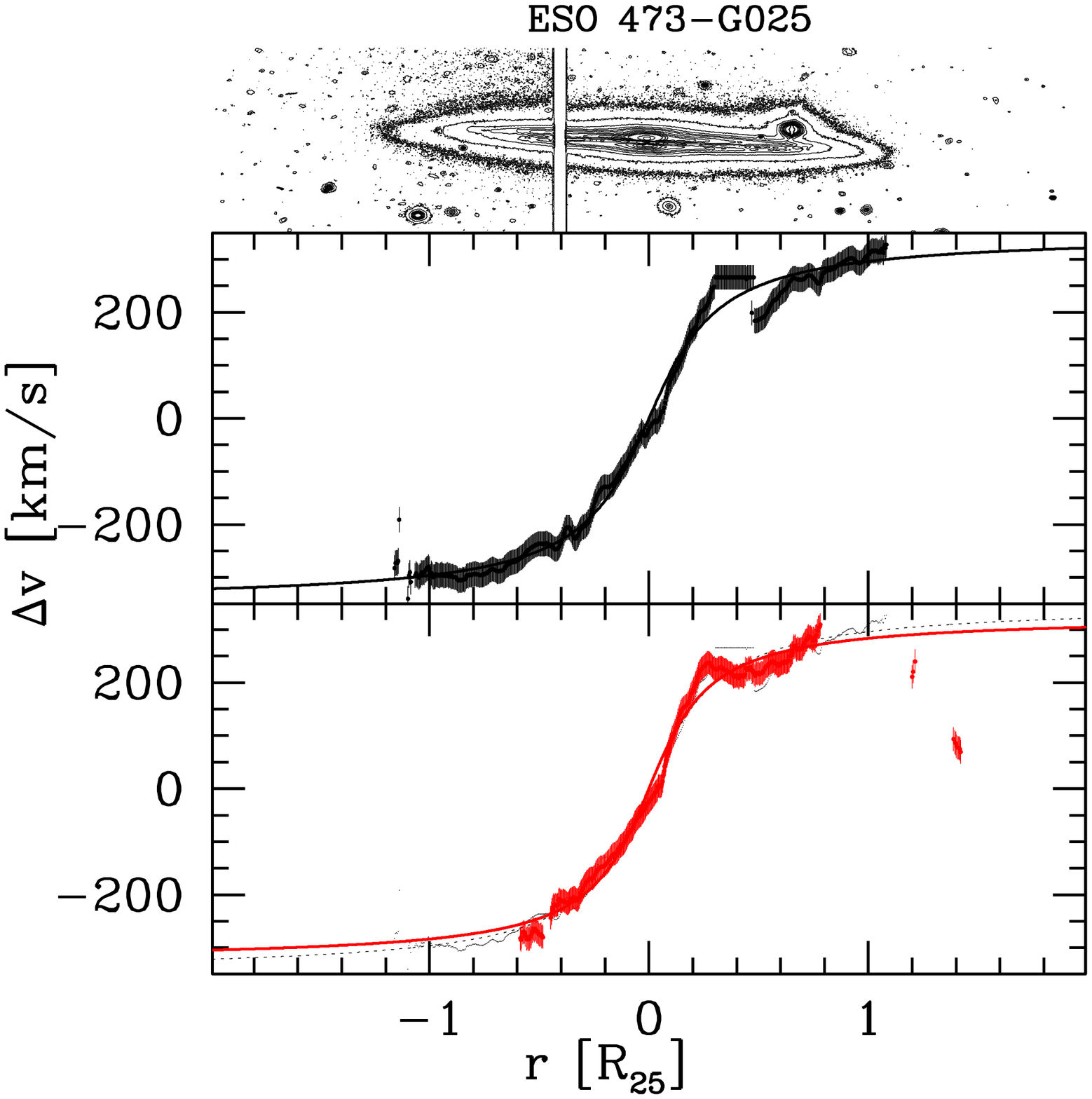}
\caption{ESO 473-G025. See text for explanation.}
\label{fig_eso473}
\end{minipage}
\end{figure}


\item {\bf ESO 473-G025} (Fig. \ref{fig_eso473}), the brightest object in our sample ($M=-20.3$), 
is an Sb-type galaxy at cz=7110 km/s. Acquisition images show what is at
most a very subtle warp. The major axis spectrum extends a little
beyond R$_{25}$. The kink visible in the rotation curve in
Fig.\ref{fig_eso473} corresponds to a region where two components of the H$\alpha$
line are visible at the same position in projection along the slit. However, there
are no signs of kinematic anomalies towards the edge of the disk. The
offset rotation curve is much less extended. A single emission region
around 1.4 R$_{25}$ coincides with what appears to be a small
satellite galaxy on the acquisition image.

\item
{\bf ESO 340-G026} (Fig. \ref{fig_eso340}) is an Sc-type galaxy at cz=5483 km/s with
$M=-19.5$. There are low-surface-brightness tails at the edge of the
disk, but they show no significant warp. We exposed in two different
offset positions, but obtained sufficient data for an analysis in only
one case. The major axis spectrum, in any case, extends to 1.3
R$_{25}$ and thus far enough to sample the kinematics of the outlying
features. The rotation curve, again, shows no signs of kinematic
anomalies.
\end{itemize}

\begin{figure}
\begin{minipage}[t]{0.5\linewidth} 
\centering
\includegraphics[width=6cm]{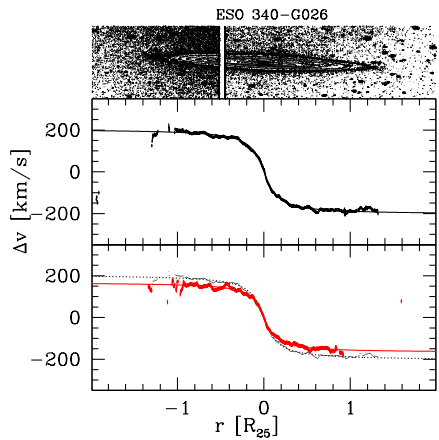}
\caption{ESO 340-G026. See text for explanation.}
\label{fig_eso340}
\end{minipage}
\end{figure}

\section{Conclusions}

In an effort to measure the kinematics of gas in the warps in the
outer disks of galaxies with high spatial resolution, we have
targetted seven edge-on galaxies, most of them with known optical
warps, using multi-hour long slit spectroscopy.

\begin{itemize}
\item In at least three of seven cases, we have observed more extended
H$\alpha$ emission if the slit was aligned along the warp rather than
along the major axis. This shows that we have been successful at
detecting gas in the warps, and that we have thus attained our
observational goal. Counting the single emission region detected in
UGC 12423, this ratio is four out of seven.
\item In three of these four cases, there are no signs of kinematic
anomalies in the rotation curve associated with the onset of the warp.
\item In one of these four cases, there is a sharp break in the
rotational velocity roughly at the onset of the warp. In this case,
the optical appearance of the ``warp'' is highly suggestive of a tidal
feature. However, the break occurs in the major axis rotation curve.
\item In the remaining objects, the outer-disk features are sampled by
the major-axis spectrum, but there is no evidence for kinematic
anomalies in these regions.
\end{itemize}

Based on a small data set (four kinematic detections of extraplanar
outer disk gas likely to be associated with a warp), we find evidence
for kinematic anomalies only in the one case whose optical appearance
strongly suggests a tidal feature. In other cases, the kinematics of
the presumed warp gas are consistent with the extrapolation of the
inner disk rotation curve. This sample is therefore not in support of the
hypothesis that the outer, warped disks are in direct contact and
exchanging angular momentum with material being newly
accreted; however, the possibility cannot be ruled out yet that the real point of
contact between the galaxy disk and accreted material lies at larger
radii, and that the gas that we see at the onset of the optical warp
has already settled into disk-like kinematics. We propose to explore
three avenues for further investigation: (i) Gas at larger radii,
substantially beyond the optical warps, should be targetted; ideally,
targets should thus be selected on the basis of HI maps where
available. (ii) The sample should be increased in size from the present
three clear detections. (iii) Additional indicators that might
constrain the origin of the warp gas, in particular, metallicity,
should be included in the analysis.

Based on observations made with ESO Telescopes at the Paranal Observatories under programme ID $<$079.B-0426$>$.


\begin{thebibliography}{}

\bibitem[Binney (1991)]{binney91}
{Binney, J.} 1992, 
 \textit{ARA\&A} ),30, 51

\bibitem[Binney, Jiang \& Dutta (1998)]{bjd}
{Binney, J., Jiang, I.-G., Dutta, S.} 1998,
\textit{MNRAS} 297, 1237

\bibitem[Bland-Hawthorn, Freeman \& Quinn (1997)]{bfq}
{Bland-Hawthorn, J., Freeman, K. C., Quinn, P. J.} 1997, 
 \textit{ApJ} ), 490, 143
\bibitem[Bosma (1981)]{bosma}
{Bosma, A.} 1981, 
\textit{AJ}, 86, 1791

\bibitem[Briggs (1990)]{briggs}
{Briggs, F. H.} 1990, 
\textit{ApJ}, 352, 15

\bibitem[Christlein \& Zaritsky (2008)]{cz}
{Christlein, D., Zaritsky, D.} 2008, 
 \textit{ApJ} ), 680, 1053

\bibitem[Cox et al. (1996)]{cox}
{Cox, A. L., Sparke, L. S., van Moorsel, G., Shaw, M.} 1996, 
\textit{AJ}, 111, 1505

\bibitem[Dekel \& Shlosman (1983)]{dekel}
{Dekel, A., Shlosman, I,} 1983, 
\textit{IAUS}, 100, 187

\bibitem[Dubinksi \& Kuijken (1995)]{dubinski}
{Dubinski, J., Kuijken, K.} 1995, 
\textit{ApJ}, 442, 492

\bibitem[Hunter \& Toomre (1969)]{huntertoomre}
{Hunter, C., Toomre, A.} 1969, 
\textit{ApJ}, 155, 747

\bibitem[Jiang \& Binney (1999)]{jiang}
{Jiang, I.-G., Binney, J.} 1999, 
 \textit{MNRAS} ), 303, 7

\bibitem[Kuijken (1991)]{kuijken}
{Kuijken, K.} 1991, 
\textit{ApJ}, 376, 467

\bibitem[Lynden-Bell (1965)]{lyndenbell}
{Lynden-Bell, D.} 1965, 
\textit{MNRAS}, 129, 299

\bibitem[Nelson \& Tremaine (1999)]{nelson}
{Nelson, R. W., Tremaine, S.} 1999, 
\textit{MNRAS}, 306, 1

\bibitem[New et al. (1998)]{new}
{New, K. C. B., Tohline, J. e., Frank, J., Vaeth, H. M.} 1998, 
 \textit{ApJ}, 503, 632

\bibitem[Ostriker \& Binney (1989)]{ostriker}
{Ostriker, E. C., Binney, J.} 1989,
 \textit{MNRAS}, 237, 785

\bibitem[Revaz \& Pfenniger (2004)]{revaz}
{Revaz, Y., Pfenniger, D.} 2004,
\textit{A\&A}, 425, 67 

\bibitem[Sancisi (1976)]{sancisi}
{Sancisi, R.} 1976,
\textit{A\&A}, 53, 159

\bibitem[Sanchez-Saavedra et al. (1990)]{sanchezsaavedra}
{Sanchez-Saavedra, M. L., Battaner, E., Florido, E.} 1990, 
\textit{MNRAS}, 246, 458

\bibitem[Sanchez-Saavedra et al. (2003)]{sanchezsaavedra2}
{Sanchez-Saavedra, M. L., Battaner, E., Guijarro, A.,
L\'opez-Correidora, M., Castro-Rodr\'iguez, N} 2003, 
\textit{A\&A}, 399, 457

\bibitem[Shen \& Sellwood (2006)]{shen}
{Shen, J., Sellwood, J. A.} 2006, 
 \textit{MNRAS} ), 370, 2

\bibitem[Sparke \& Casertano (1988)]{sparke}
{Sparke, L. S., Casertano, S.} 1988,
\textit{MNRAS}, 234, 873

\bibitem[Toomre (1983)]{toomre}
{Toomre, A.} 1983, 
\textit{IAUS}, 100, 177

\end{thebibliography}
\end{document}